\documentclass[english,prb, twocolumn, superscriptaddress]{revtex4-1}
\usepackage[T1]{fontenc}
\usepackage[latin9]{inputenc}
\usepackage{verbatim}
\usepackage{textcomp}
\usepackage{amstext}
\usepackage{graphicx}

\makeatletter
\@ifundefined{textcolor}{}
{%
 \definecolor{BLACK}{gray}{0}
 \definecolor{WHITE}{gray}{1}
 \definecolor{RED}{rgb}{1,0,0}
 \definecolor{GREEN}{rgb}{0,1,0}
 \definecolor{BLUE}{rgb}{0,0,1}
 \definecolor{CYAN}{cmyk}{1,0,0,0}
 \definecolor{MAGENTA}{cmyk}{0,1,0,0}
 \definecolor{YELLOW}{cmyk}{0,0,1,0}
 }

\usepackage{graphicx}
\usepackage{caption}
\DeclareCaptionLabelFormat{natfig}{Figure #2}
\DeclareCaptionLabelFormat{nattable}{Table #2}
\DeclareCaptionLabelSeparator{pipe}{ | }
\makeatother

\usepackage{babel}
\begin{document}

\title{From the SrTiO$_{3}$ surface to the LaAlO$_{3}$/SrTiO$_{3}$ interface:
How thickness is critical}

\author{N.~C.~Plumb}

\email{nicholas.plumb@psi.ch}

\affiliation{Swiss Light Source, Paul Scherrer Institut, CH-5232 Villigen PSI,
Switzerland}

\author{M.~Kobayashi}

\affiliation{Swiss Light Source, Paul Scherrer Institut, CH-5232 Villigen PSI,
Switzerland}

\affiliation{Department of Applied Chemistry, School of Engineering, University
of Tokyo, 7-3-1 }

\author{M.~Salluzzo}

\affiliation{CNR-SPIN, Complesso Universitario Monte S. Angelo, Via Cinthia I-80126
Napoli, Italy}

\author{E.~Razzoli}

\author{C.~E.~Matt}

\author{V.~N.~Strocov}

\author{K.-J.~Zhou}

\author{C.~Monney}

\author{T.~Schmitt}

\author{M.~Shi}

\affiliation{Swiss Light Source, Paul Scherrer Institut, CH-5232 Villigen PSI,
Switzerland}

\author{J.~Mesot}

\affiliation{Swiss Light Source, Paul Scherrer Institut, CH-5232 Villigen PSI,
Switzerland}

\affiliation{Institut de la Matiere Complexe, EPF Lausanne, CH-1015, Lausanne,
Switzerland}

\author{L.~Patthey}

\affiliation{Swiss Light Source, Paul Scherrer Institut, CH-5232 Villigen PSI,
Switzerland}

\affiliation{SwissFEL, Paul Scherrer Insitut, CH-5232 Villigen PSI, Switzerland}

\author{M.~Radovic}

\email{milan.radovic@psi.ch}

\affiliation{Swiss Light Source, Paul Scherrer Institut, CH-5232 Villigen PSI,
Switzerland}

\affiliation{Institut de la Matiere Complexe, EPF Lausanne, CH-1015, Lausanne,
Switzerland}

\affiliation{SwissFEL, Paul Scherrer Insitut, CH-5232 Villigen PSI, Switzerland}
\begin{abstract}
Novel properties arising at interfaces between transition metal oxides,
particularly the conductivity at the interface of LaAlO$_{3}$ (LAO)
and SrTiO$_{3}$ (STO) band insulators, have generated new paradigms,
challenges, and opportunities in condensed matter physics. Conventional
transport measurements have established that intrinsic conductivity
appears in LAO/STO interfaces when the LAO film matches or exceeds
a critical thickness of 4 unit cells (uc). Recently, %
a number of experiments raise important questions about the role of
the LAO film, the influence of photons, and the effective differences
between vacuum/STO and LAO/STO, both above and below the standard
critical thickness. Here, using angle-resolved photoemission spectroscopy
(ARPES) on \emph{in situ} prepared samples, as well as resonant inelastic
x-ray scattering (RIXS), we study how the metallic STO surface state
evolves during the growth of a crystalline LAO film. In all the samples,
the character of the conduction bands, their carrier densities, the
Ti$^{3+}$ crystal fields, and the responses to photon irradiation
bear strong similarities. However, LAO/STO interfaces exhibit intrinsic
instability toward in-plane%
{} folding of the Fermi surface at and above the 4-uc thickness threshold.
This ordering distinguishes these heterostructures from bare STO and
sub-critical-thickness LAO/STO and coincides with the onset of unique
properties such as magnetism and built-in conductivity.
\end{abstract}
\maketitle
Unraveling the origins of the electron gases at the STO surface\cite{Santander-Syro2011,Meevasana2011,DiCapua2012}
and LAO/STO interfaces\cite{Ohtomo2004}, as well as the mechanisms
leading to interesting and useful behaviors such as magnetism\cite{Brinkman2007,Kalisky2012,Salman2012}
and persistent photo-enhanced/-induced conductivity\cite{Ristic2012,diUccio2012},
are crucial steps towards fully exploiting the technological promise\cite{Mannhart2010}
of these systems. %
Theories about the existence of the metallic interface in LAO/STO
have largely centered around growth-induced defects and doping such
as ion intermixing\cite{Willmott2007} and oxygen vacancies, as well
as the \textquotedblleft{}polar catastrophe\textquotedblright{} model
in which, at the critical thickness of 4 uc\cite{Thiel2006}, carriers
transfer across LAO to the STO interface in order to quench the electrostatic
potential arising from the growth of polar LAO on nonpolar STO\cite{Nakagawa2006}.
While several arguments can be made in favor of the polar catastrophe
model\cite{Schlom2011}, the issue is not yet settled. %
Meanwhile the origin of interfacial magnetism in LAO/STO remains an
open question as well.

A number of recent observations enrich the story surrounding the origins
of conductivity at oxide interfaces and surfaces. For instance, interfacial
conducting states have been found in other STO-based heterostructures
where no polar discontinuity should occur\cite{Annadi2012a,Chen2011b,Chen2013,Herranz2012}.
These results come amidst the backdrop of the discovery of a metallic
state at the surface of bare STO. %
Such observations draw attention to the role of the STO substrate
and raise the possibility that STO\textquoteright{}s surface and interface
gases are intimately connected; therefore each may be able to teach
us about the origins of metallicity and other properties in the other.
With this in mind, we set out to investigate the evolution of the
low-dimensional metallic state on STO starting from the bare surface
and progressing through 4 or more unit cells of LAO over STO. 

\begin{figure*}
\includegraphics[width=0.7\textwidth]{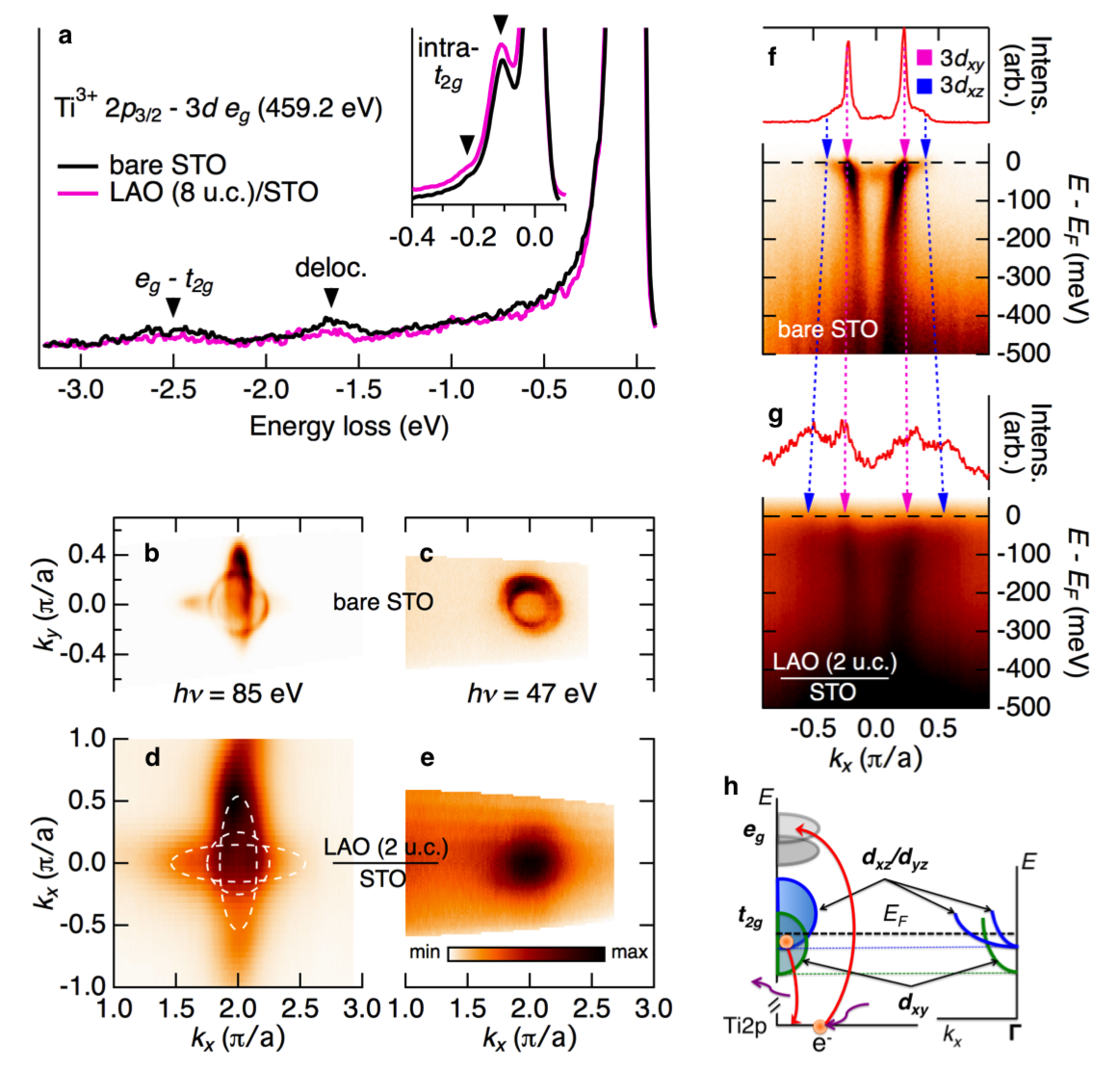}

\caption{\textbf{\label{fig:fig1}Comparisons of bare STO and LAO/STO. a,}
RIXS spectra of STO and LAO/STO. The LAO layer is 8 uc  thick. The
$e_{g}-t_{2g}$e, delocalized electronic, and intra-$t_{2g}$ transitions
(inset) are highlighted. \textbf{b--e,} Conventional ARPES measurements
of the Fermi surface of STO under 2 uc  of LAO and bare STO  at
47 eV and 85 eV in the second Brillouin zone. \textbf{d, e,} Dispersion
cuts in the first Brillouin zone through at $k_{y}=0$ for bare STO
and STO with 2 unit cells of LAO deposited on top. The photon energy
was 85 eV. The line profiles show the spectral weight at $E_{F}$,
and the markers indicate the Fermi momenta of the outer $3d_{xy}$
ring (magenta) and $3d_{xz}$ (blue) bands. The color scales have
been enhanced to emphasize key features.  The sketched Fermi surfaces
in \textbf{d} are based on the Fermi momenta as judged from \textbf{e}.
\textbf{f,} Schematic views of the intra-$t_{2g}$ RIXS process and
the corresponding band structure which agrees with the ARPES data.}
\end{figure*}

We begin our study using RIXS to probe the crystal field excitations
in bare STO and a sample with 8 uc of LAO grown over STO. As a \textquotedblleft{}photon
in, photon out\textquotedblright{} technique, RIXS penetrates into
the bulk of the sample to probe electron transitions between unoccupied
states. %
By performing RIXS on resonance with Ti$^{3+}$ absorption features,
the technique can be tuned to focus on the surface and interface regions
in STO and LAO/STO, respectively, where the low-dimensional electron
gases reside\cite{DiCapua2012,Plumb2013,Sing2009,Cancellieri2013,Zhou2011}.
Figure \ref{fig:fig1}a compares RIXS spectra from a bare TiO$_{2}$-terminated
STO substrate and one with an LAO film grown on top (8 uc). %
In order to selectively probe the conducting states, we performed
RIXS using $h\nu=459.2$ eV, corresponding to $2p_{3/2}\text{--}3d\: e_{g}$
absorption by Ti$^{3+}$ sites. Both samples show a peak near the
elastic line at an energy loss of -110 meV and a weak shoulder at
roughly -200 meV, as well as features centered at approximately -2.5
eV and -1.7 eV. Based on our previous RIXS study of LAO/STO multilayers\cite{Zhou2011},
we attribute the near-elastic features as intra-$t_{2g}$ transitions,
the structure at -2.5 eV as $e_{g}\text{--}t_{2g}$ transitions, and
the one at -1.7 eV as arising from delocalized electron excitations.
Overall, the spectra clearly demonstrate that cubic symmetry is broken\cite{Zhou2011},
and the correspondence between these spectral features in both bare
STO and LAO/STO is a clear indication that Ti$^{3+}$ sites in both
systems are associated with essentially the same structural distortion
\textendash{} most likely the polar Ti-O buckling seen by several
studies of LAO/STO interfaces\cite{Pauli2011,Cantoni2012} and the
bare STO surface\cite{Bickel1989,Hikita1993,Ikeda1999}.

Next we studied STO perturbed under 2 uc of LAO film using \emph{in
situ} ARPES at conventional UV photon energies. Even though this thickness
of LAO/STO %
does not have a conducting interface in ambient, dark conditions,
scanning tunneling %
measurements have successfully been performed on such samples when
illuminated, implying the existence of conductivity\cite{Ristic2012}.
Moreover, a Fermi surface appears when the sample is %
irradiated in ultrahigh vacuum (UHV, $\sim10^{-11}$ mbar). This behavior
is similar to how the electron gas found on bare STO can form under
analogous conditions\cite{Meevasana2011,Plumb2013}. Fermi surfaces
of the bare STO and 2 uc LAO/STO systems obtained using photon energies
of 85 eV and 47 eV (corresponding to different cuts through the out-of-plane
momentum axis $k_{z}$\cite{Plumb2013}) are shown in Figs. \ref{fig:fig1}b-e.
Figs. \ref{fig:fig1}f-g show side-by-side comparisons of energy-vs.-$k_{x}$
dispersion data from bare STO and 2 uc LAO/STO obtained at $k_{y}=0$
in the first Brillouin zone using $h\nu=85$ eV. The ellipsoidal components
(heavy bands) of the FS extend slightly farther in $k$ than in bare
STO (a little over 0.5 $\pi/a$ along the long axes from the Brillouin
zone center, instead of $\sim0.4$ $\pi/a$ for bare STO\cite{Plumb2013}),
qualitatively consistent with calculations\cite{Popovic2008}. The
overall nature of the near-$E_{F}$ electronic structure of the LAO/STO
system is %
similar to the bare surface of STO, whose photon energy dependence
in ARPES %
is attributed to a combination of confined, non-bulk-like quasi-2D
and 3D electronic states associated with Ti $3d_{xy}$ and Ti $3d_{xz}$/$3d_{yz}$
bands, respectively\cite{Plumb2013}. This naturally suggests that
the conducting $3d_{xy}$ and $3d_{xz}$/$3d_{yz}$ electrons comprising
the photo-induced metal in the 2 uc LAO/STO interface region have
similar spatial distributions to those in STO. Qualitatively, this
result is in agreement with several experimental\cite{Copie2009,Dubroka2010}
and theoretical\cite{Popovic2008,Delugas2011,Stengel2011} studies
of thicker ($\geq4$ uc) LAO/STO interfaces, which find that the electron
gas actually extends several unit cells into STO. Moreover, the distinct
dimensionalities and effective mass characteristics of the $3d_{xy}$
and $3d_{xz}$/$3d_{yz}$ electrons naturally explain the observation
of two types of carriers in many experiments\cite{Zhou2011,Brinkman2007,Savoia2009,BenShalom2010}.

Studying the electronic structure of buried interface states presents
a significant challenge for photoemission experiments, since the inelastic
mean free path %
of photoelectrons in a solid is generally considered to be under 1
nm over a broad range of kinetic energies%
. %
Thus we continued \emph{in situ} ARPES measurements for samples with
4 and 5 uc of LAO by using higher photon energies where resonance
enhancement at the Ti $L$ edge occurs, %
thereby allowing us to view the momentum-resolved electronic structure
associated with Ti.

\begin{figure*}
\includegraphics[width=0.7\textwidth]{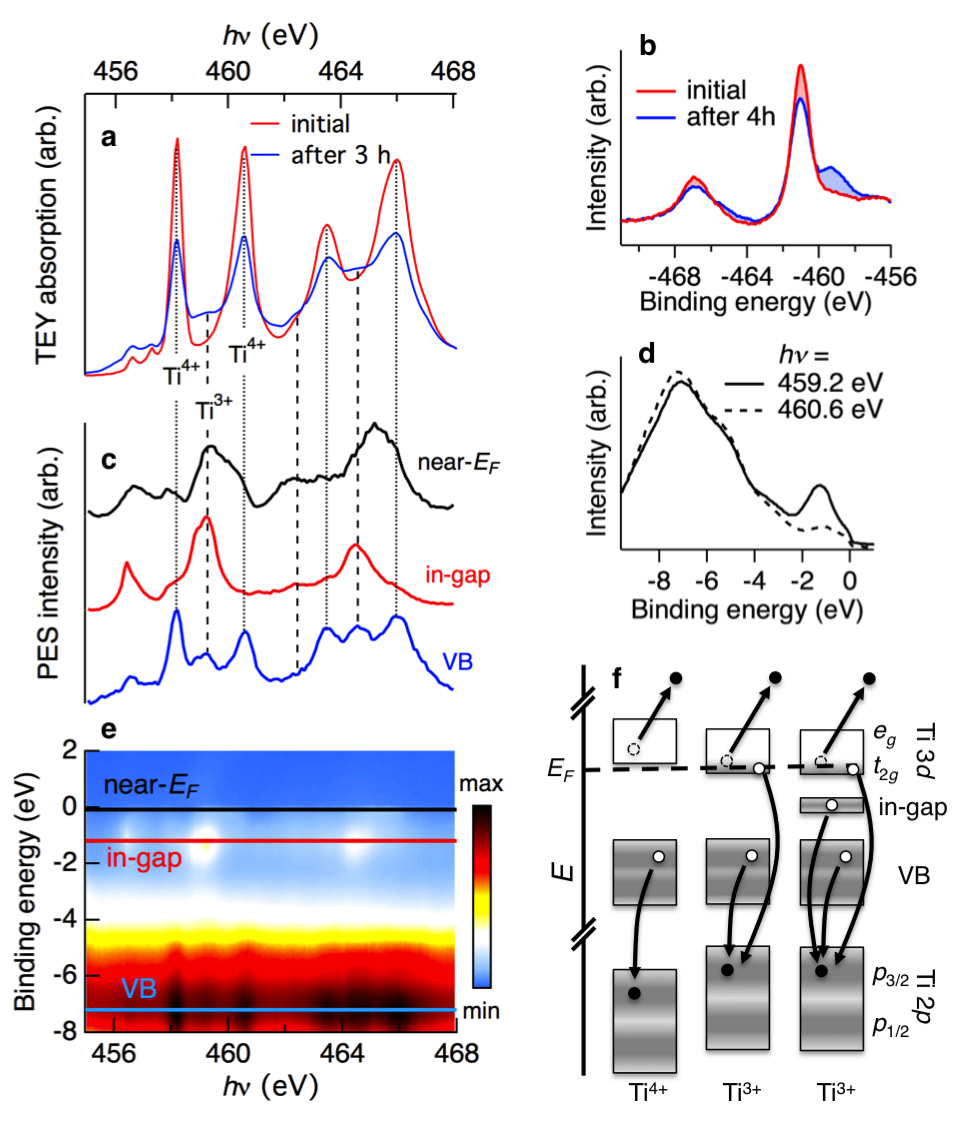}

\caption{\textbf{\label{fig:fig2}Resonant photoemission spectroscopy of the
interface of 5 uc LAO grown over STO. a,} Absorption spectra measured
by total electron yield (TEY) at an initial time and after 3 hours
of exposure to synchrotron radiation under the experimental conditions.
\textbf{b,} Ti $2p$ XPS measured at an initial time and after 4 hours
of irradiation. \textbf{c,} Cuts the of angle-integrated PES intensity
vs. $h\nu$ at binding energies of -0.1 eV (near-$E_{F}$), -1.3 eV
(in-gap), and in the valence band at -7.1 eV (VB). \textbf{d,} Comparison
of angle-integrated PES data at $h\nu=460.6$ eV (a Ti$^{4+}$ resonance)
and $h\nu=459.2$ eV (a Ti$^{3+}$ resonance). \textbf{e,} Map of
the PES signal as a function of the incoming photon energy. \textbf{f,}
Schematic of resonant photoemission decay channels in the Ti$^{4+}$
and Ti$^{3+}$ environments.}

\end{figure*}

Figure \ref{fig:fig2} presents results from resonant angle-integrated
photoemission spectroscopy (resonant PES), absorption measured by
the total electron yield (TEY), and core level x-ray photoemission
spectroscopy (XPS) performed on 5 uc of LAO grown over STO. The sample
underwent post-growth annealing in oxygen ($P_{O_{2}}=200$ mbar,
$T=55$ $^{\circ}$C, 1 h). Over the course of the measurement, exposure
to synchrotron radiation causes the spectral characteristics of the
LAO/STO sample to evolve in a manner indicating that the metallicity
at the interface becomes enhanced. In particular, we observe a transfer
of spectral weight within the TEY absorption from $ $Ti$^{4+}$ to
Ti$^{3+}$ peaks (Fig.~\ref{fig:fig2}a), concurrent with a Ti$^{4+}$-to-Ti$^{3+}$
redistribution of counts in the Ti $2p$ XPS doublet (Fig.~\ref{fig:fig2}b).
Complimentary %
resonant PES data (Fig.~\ref{fig:fig2}c--e) were obtained by tuning
the photon energy over the full Ti $2p\text{--}3d$ absorption range.
The high intensity at the deepest binding energy range of Fig.~\ref{fig:fig2}e
comes from the O $2p$ valence band, %
which resonates on both Ti$^{3+}$ and Ti$^{4+}$ energies. This demonstrates
a noticeable degree of covalency/hybridization in the Ti-O bonds,
which may be relevant to theorectical descriptions of the LAO/STO
interface. The features that appear at a binding energy of approximately
-1.3 eV come from the ``in-gap'' states previously observed in the
case of both STO single crystals\cite{Meevasana2011,Plumb2013} and
LAO/STO heterostructures\cite{Koitzsch2011a,Ristic2012,Drera2011},
while the spectral weight at the Fermi level indicates that metallicity
is present in the sample. Interestingly, the resonance energies of
the in-gap states match well with the peak structure of Ti$^{3+}$
(known from LaTiO$_{3}$\cite{Abbate1991,Salluzzo2009}), whereas
states at the Fermi edge resonate slightly off these energies. We
speculate that this slight discrepancy might indicate the existence
of different varieties of nominal Ti$^{3+}$ --- perhaps located at
different depths in the near-interface region and/or with different
effective valence Ti$^{3+\delta(z)}$. Example sketches of resonant
photoemission processes corresponding with the level structures of
the Ti$^{4+}$ and Ti$^{3+}$ states are depicted in Fig.~\ref{fig:fig2}g.
The photon energy ranges where the in-gap and Fermi level states resonate
overlap well with the Ti$^{3+}$ features of the TEY absorption spectrum
that become enhanced during photon irradiation. This behavior is very
similar to the evolution seen on the surface of bare STO during illumination,
where the enhancement %
of in-gap and metallic states go hand-in-hand\cite{Meevasana2011,Plumb2013}.

\begin{figure*}
\includegraphics[width=1\textwidth]{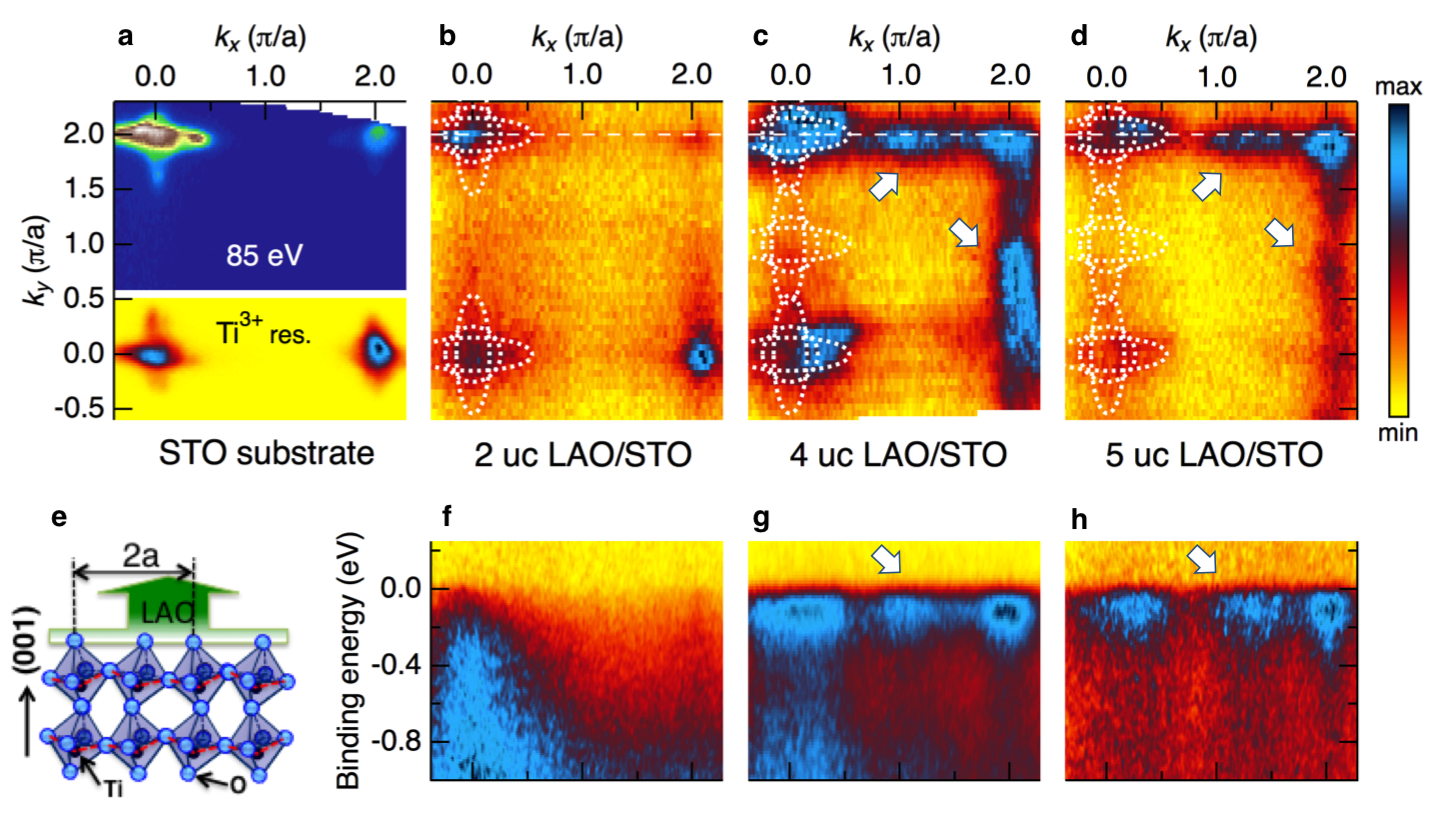}

\caption{\textbf{\label{fig:fig3}Evolution of the Fermi surface on STO during
deposition of LAO. a,} Fermi surface obtained from the surface of
a bare STO substrate. The upper map presents data in the $(k_{x},k_{y})=(0,\pi/a)$
and $(\pi/a,\pi/a)$ Brillouin zones obtained by conventional UV
ARPES at $h\nu=85$ eV. The bottom map shows analogous high energy
resonant ARPES data obtained using performed on a Ti$^{3+}$ resonance
in the $(0,0)$ and $(0,\pi/a)$ zones. \textbf{b,} Fermi surface
of the interface of 2 uc LAO deposited over STO mapped by resonant
ARPES. \textbf{c,} Interface of 4 uc LAO over STO. \textbf{d,} Interface
of 5 uc LAO over STO. The sketched Fermi surfaces are based on the
measurement of 2 uc LAO/STO in Fig.~\ref{fig:fig1}. \textbf{f--h,}
$E$-vs.-$k_{x}$ cuts along $k_{y}=2\pi/a$ for each of the above
maps (\textbf{b--d}). When LAO is deposited above 4 uc, signs of an
in-plane reconstruction, marked with arrows, are visible. \textbf{e,}
Schematic view of TiO$_{6}$ octahedral rotations that give rise
to $2\times1$ ordering in the $x$-$y$ plane. We probed multiple
Ti$^{3+}$ resonances to confirm the findings; shown here, e.g., are
$h\nu=465.3$ eV (\textbf{a}), 462.6 eV (\textbf{b}, \textbf{d}),
and 459.2 eV (\textbf{c}).}

\end{figure*}

At this point the similarities between the metallic states seen on
the surface of irradiated bare STO and at the LAO/STO interface ---
whether below (2 uc of LAO) or above the standard critical thickness
(4--8 uc of LAO) --- are striking. All exhibit similar electronic
structure, Ti$^{3+}$ crystal fields, correspondence between the existence
of in-gap and metallic electronic states, and creation/enhancement
of such states via irradiation. However, the Fermi surfaces of LAO/STO
where LAO is at least 4 uc thick show a new characteristic not found
in thinner samples or on bare STO. Figure \ref{fig:fig3} depicts
\emph{in situ} resonant ARPES measurements on bare STO and LAO/STO
bilayers with LAO thicknesses of 2, 4, and 5 uc. The data were acquired
on resonance with Ti$^{3+}$ absorption features. Figure \ref{fig:fig3}a
shows Fermi surface mapping on bare STO using resonant and conventional
ARPES%
.%

\begin{figure}
\includegraphics[width=1\columnwidth]{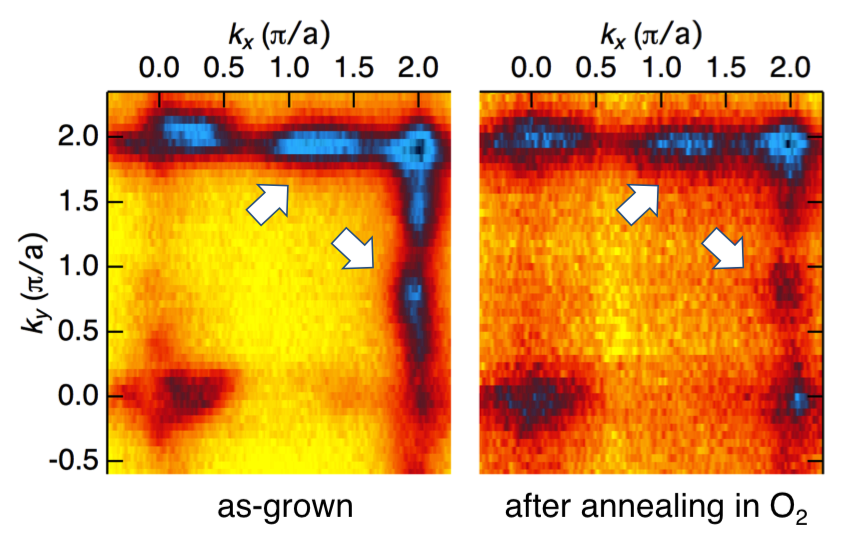}

\caption{\label{fig:fig4}\textbf{Comparison of Fermi surfaces of LAO/STO before
(left) and after (right) annealing in oxygen atmosphere.} The data
were obtained using $h\nu=$465.3 eV. The arrows point to signatures
of an in-plane reconstruction. }

\end{figure}

Putting aside %
fine details concerning the differences between the Fermi surfaces
seen by conventional and resonant ARPES, we notice a substantive change
occurring at the standard critical LAO thickness. Although all the
studied samples show $1\times1$ ordering of the LAO surfaces (see
Supplementary Information), in those where the thickness of LAO is
at least 4 uc (Fig.~\ref{fig:fig3}c--d), the Fermi surface of the
interface shows folding along the $k_{x}$ and $k_{y}$ directions
--- possibly of a twinned $2\times1$ order. %
Such folding is not seen for either bare STO or 2 uc LAO/STO (Fig.~\ref{fig:fig3}a--b).
This can also been seen by dispersion cuts along $k_{y}=2\pi/a$ in
the second Brillouin zone for each of the Fermi surface maps (Fig.~\ref{fig:fig3}f--h).
We observed the same folding pattern (Fig.~\ref{fig:fig4}) after
annealing the 5 uc LAO/STO sample \emph{in situ} under in high oxygen
pressure (200 mbar at 500 \textdegree{}C for 1 h). This strongly suggests
that the in-plane %
ordering is not directly associated with oxygen vacancies or other
defects. With this in mind, the simplest interpretation is that a
$2\times1$ folding occurs due to twinned rotations of the TiO$_{6}$
octahedra about the {[}100{]}/{[}010{]} axes. The proposed rotation
pattern is sketched in Fig.~\ref{fig:fig3}e, which also qualitatively
depicts the ferroelectric Ti-O buckling distortion. 

The occurrence of octahedral tilting about {[}100{]}/{[}010{]} or
a similar in-plane reconstruction in sufficiently thick LAO/STO interfaces
may be connected to unique properties arising in these systems. %
In particular, various tilting distortions (sometimes combined with
ferroelectric Ti-O buckling) in LAO/STO interfaces might correspond
with different charge- and/or magnetically-ordered phases, which may
even compete in the ground state\cite{Zhong2008}. Indeed, the onset
of the in-plane %
ordering seen here after depositing 4 or more uc of LAO coincides
with the appearance of dilute magnetism in samples studied by scanning
superconducting quantum interference device microscopy\cite{Kalisky2012}.
Similarly, LAO/STO superlattices were found to exhibit magnetic behavior
only when the LAO spacer layers have a minimum thickness of $\sim4\text{--}6$
uc\cite{Salman2012}. 

Concerning the origin of built-in conductivity in LAO/STO with at
least 4 uc of LAO, it has been shown that the LAO core levels do not
exhibit strong thickness dependent shifts or broadening, as would
be expected on the basis of the polar catastrophe model\cite{Segal2009,Slooten2013}.
Furthermore, hole pockets from LAO that should occur at the Brillouin
zone corners are absent in our ARPES data, which is in agreement with
an %
\emph{ex situ} study\cite{Berner2013}. Since the experiments here
are performed on samples not exposed to the atmosphere, arguments
in favor of an extrinsic passivation of the surface to explain this
discrepancy become less likely. Our results also do not favor a direct
connection between oxygen vacancies and the built-in conductivity
in LAO/STO. The 5 uc LAO/STO sample shows little qualitative change
before and after oxygen annealing (Fig.~\ref{fig:fig4}), consistent
with the idea that oxygen vacancies, while perhaps providing some
additional doping, do little to fundamentally alter the interface
electron gas\cite{Basletic2008}. %
The electronic structure and doping of the metallic state on the surface
of bare STO similarly shows no direct link to the concentration of
oxygen vacancies\cite{Santander-Syro2011,Plumb2013}. 

While the photoemission experiments do not provide specific confirmation
of the origin of the carriers at the interface, they make it clear
that distortions in the surface/interface region of STO strongly influence
the electronic structure. The polar, ferroelectric-type Ti-O buckling
distortion frequently reported on the STO surface and in the interfacial
region of STO with LAO is generally thought to play a key role in
the spatial confinement of the conducting electrons and compensating
the polarity from the LAO %
{} film \cite{Schwingenschlogl2009,Stengel2011}. Being common to all
these systems, Ti-O buckling is likely to account for the shared signatures
of symmetry-breaking in the electronic structure: namely the split
$3d_{xy}\text{--}3d_{xz}/3d_{yz}$ conduction bands with indications
of both 2D and 3D-like characteristics and the non-cubic crystal field
seen in Ti$^{3+}$ RIXS spectra. The polar buckling region should
naturally accommodate carriers\cite{Stengel2011,Plumb2013}, and STO
would therefore seem to be a particularly suitable host for low-dimensional
electron gases in large part due to its innate propensity for forming
this structure near its surface/interface. 

It is interesting to consider how $2\times1$ or similar in-plane
ordering might be connected to the Ti-O buckling and how, in turn,
such ordering might influence the conducting properties of LAO/STO
interfaces. Recently, theoretical work has explored the concept of
``improper ferroelectricity'' in the context of oxide heterostructures,
in which ferroelectric displacements occur via coupling to a second
order parameter \textendash{} e.g., various forms of octahedral tilting
which are not by themselves polar\cite{Stengel2012}. In principle,
even some tilting patterns that in bulk would normally compete with
ferroelectric distortions can in certain cases give rise to improper
ferroelectricity when the symmetry of the system is broken by an interface\cite{Stengel2012,Bousquet2008}.
Moreover, it has been proposed that the walls between domains of tilted
octahedra (e.g., $2\times1$ twin boundaries) in STO may induce polarization\cite{Morozovska2012}.
Hence the in-plane %
ordering might strongly influence the nature and/or strength of polar
distortions near the interface, which in turn are closely linked to
the extent of the confinement region and thus the transport properties. 

Multiple effects (e.g., defects, photons, strain) may trigger or otherwise
alter the surface/interface structure, thereby affecting the conductivity
or other properties. The role of photons in inducing/enhancing the
surface and interface conductivity of STO and LAO/STO, respectively,
is particularly %
intriguing. The appearance and/or increase of spectral weight at $E_{F}$
and the enhanced signatures of Ti$^{3+}$ in the core level and absorption
spectra appear to be long-lived; we can move the synchrotron beam
away from a heavily exposed spot for hours, then go back, and find
the spectral changes largely intact. This is consistent with the persistent
photo-induced changes seen in the electronic structure of the STO
surface\cite{Plumb2013} and in the transport behavior of LAO/STO
and related interfaces\cite{diUccio2012}. The extremely long %
lifetimes of the changes imply that photons do more than merely generate
carriers by exciting electrons to the conduction band. We therefore
propose that light can directly and/or indirectly play some role in
altering the structure of the surface/interface region of STO in terms
of the polar Ti-O buckling\cite{Plumb2013}, $2\times1$ or other
ordering, and/or the domain structure \textendash{} all of which could
impact the conductivity and spectroscopic properties %
as discussed above. 

Whatever the origin of the carriers themselves on bare STO and in
LAO/STO, the results here emphasize the important role of the near-surface/-interface
structure of STO for the realization of a confined electron gas. Indeed,
if there is one unifying characteristic of the STO surface and LAO/STO
interfaces, it is that conductivity appears to be strongly linked
to the existence of ferroelectric Ti-O buckling in the surface/interface
region of STO\cite{Plumb2013,Pauli2011,Stengel2011,Cantoni2012,Salluzzo2013}.
The similar Ti$^{3+}$ crystal fields and near-$E_{F}$ electronic
structure of bare STO and LAO/STO illustrate the intimate connection
between these systems owing to the distortions in their metallic regions.
These commonalities, however, belie important differences between
the various STO-based systems, and the in-plane%
{} ordering seen in the 4- and 5-uc thick LAO/STO interfaces may be
an important new insight into how properties such as built-in conductivity
and weak/dilute magnetism emerge %
above the 4-uc threshold.

\subsection*{Acknowledgments}

The resonant ARPES and RIXS experiments were performed at the Advanced
Resonant Spectroscopies (ADRESS) beamline of the Swiss Light Source
(SLS) at the Paul Scherrer Institut in Villigen, Switzerland. The
SAXES instrument used for the RIXS experiment was jointly built by
the Paul Scherrer Institut and Politecnico di Milano, Italy. The high-resolution
ARPES measurements were carried out at the Surface/Interface Spectroscopy
(SIS) beamline of the SLS. We are grateful for especially valuable
conversations with H.~Dil and F.~Miletto-Granozio. We also shared
informative discussions with R.~Claessen, M.~Sing, and G.~Berner,
C.~Cancellieri, M.~L.~Reinle-Schmitt, and C.~W.~Schneider.

\subsection*{Author contributions}

M.~R.~conceived the project. N.~C.~P., M.~Salluzzo, E.~R., M.~K.,
C.~E.~M., C.~M., K.-J.~Z.~and M.~R.~collected data. All authors
discussed the results. N.~C.~P.~and M.~R.~wrote the paper with
valuable input from all the coauthors.

\bibliographystyle{naturemag}
\bibliography{citations}

\begin{thebibliography}{10}
\expandafter\ifx\csname url\endcsname\relax
  \def\url#1{\texttt{#1}}\fi
\expandafter\ifx\csname urlprefix\endcsname\relax\def\urlprefix{URL }\fi
\providecommand{\bibinfo}[2]{#2}
\providecommand{\eprint}[2][]{\url{#2}}

\bibitem{Santander-Syro2011}
\bibinfo{author}{Santander-Syro, A.~F.} \emph{et~al.}
\newblock \bibinfo{title}{Two-dimensional electron gas with universal subbands
  at the surface of {SrTiO$_3$}}.
\newblock \emph{\bibinfo{journal}{Nature}} \textbf{\bibinfo{volume}{469}},
  \bibinfo{pages}{189--193} (\bibinfo{year}{2011}).
\newblock \urlprefix\url{http://dx.doi.org/10.1038/nature09720}.

\bibitem{Meevasana2011}
\bibinfo{author}{Meevasana, W.} \emph{et~al.}
\newblock \bibinfo{title}{Creation and control of a two-dimensional electron
  liquid at the bare {SrTiO$_3$} surface}.
\newblock \emph{\bibinfo{journal}{Nature Mater.}}
  \textbf{\bibinfo{volume}{10}}, \bibinfo{pages}{114--118}
  (\bibinfo{year}{2011}).
\newblock \urlprefix\url{http://dx.doi.org/10.1038/nmat2943}.

\bibitem{DiCapua2012}
\bibinfo{author}{Di~Capua, R.} \emph{et~al.}
\newblock \bibinfo{title}{Observation of a two-dimensional electron gas at the
  surface of annealed {SrTiO${}_{3}$} single crystals by scanning tunneling
  spectroscopy}.
\newblock \emph{\bibinfo{journal}{Phys. Rev. B}} \textbf{\bibinfo{volume}{86}},
  \bibinfo{pages}{155425} (\bibinfo{year}{2012}).
\newblock \urlprefix\url{http://link.aps.org/doi/10.1103/PhysRevB.86.155425}.

\bibitem{Ohtomo2004}
\bibinfo{author}{Ohtomo, A.} \& \bibinfo{author}{Hwang, H.~Y.}
\newblock \bibinfo{title}{A high-mobility electron gas at the
  {LaAlO$_3$/SrTiO$_3$} heterointerface}.
\newblock \emph{\bibinfo{journal}{Nature}} \textbf{\bibinfo{volume}{427}},
  \bibinfo{pages}{423--426} (\bibinfo{year}{2004}).
\newblock \urlprefix\url{http://dx.doi.org/10.1038/nature02308}.

\bibitem{Brinkman2007}
\bibinfo{author}{Brinkman, A.} \emph{et~al.}
\newblock \bibinfo{title}{Magnetic effects at the interface between
  non-magnetic oxides}.
\newblock \emph{\bibinfo{journal}{Nature Mater.}} \textbf{\bibinfo{volume}{6}},
  \bibinfo{pages}{493--496} (\bibinfo{year}{2007}).
\newblock \urlprefix\url{http://dx.doi.org/10.1038/nmat1931}.

\bibitem{Kalisky2012}
\bibinfo{author}{Kalisky, B.} \emph{et~al.}
\newblock \bibinfo{title}{Critical thickness for ferromagnetism in
  {LaAlO$_3$/SrTiO$_3$} heterostructures}.
\newblock \emph{\bibinfo{journal}{Nat. Commun.}} \textbf{\bibinfo{volume}{3}},
  \bibinfo{pages}{922} (\bibinfo{year}{2012}).
\newblock \urlprefix\url{http://dx.doi.org/10.1038/ncomms1931}.

\bibitem{Salman2012}
\bibinfo{author}{Salman, Z.} \emph{et~al.}
\newblock \bibinfo{title}{Nature of weak magnetism in
  {${\mathrm{SrTiO}}_{3}/{\mathrm{LaAlO}}_{3}$} multilayers}.
\newblock \emph{\bibinfo{journal}{Phys. Rev. Lett.}}
  \textbf{\bibinfo{volume}{109}}, \bibinfo{pages}{257207}
  (\bibinfo{year}{2012}).
\newblock
  \urlprefix\url{http://link.aps.org/doi/10.1103/PhysRevLett.109.257207}.

\bibitem{Ristic2012}
\bibinfo{author}{Ristic, Z.} \emph{et~al.}
\newblock \bibinfo{title}{Photodoping and in-gap interface states across the
  metal-insulator transition in {LaAlO${}_{3}$/SrTiO${}_{3}$}
  heterostructures}.
\newblock \emph{\bibinfo{journal}{Phys. Rev. B}} \textbf{\bibinfo{volume}{86}},
  \bibinfo{pages}{045127} (\bibinfo{year}{2012}).
\newblock \urlprefix\url{http://link.aps.org/doi/10.1103/PhysRevB.86.045127}.

\bibitem{diUccio2012}
\bibinfo{author}{{di Uccio}, U.~S.} \emph{et~al.}
\newblock \bibinfo{title}{{Reversible and Persistent Photoconductivity at the
  NdGaO3/SrTiO3 Conducting Interface}}.
\newblock \emph{\bibinfo{journal}{ArXiv e-prints}}  (\bibinfo{year}{2012}).
\newblock \urlprefix\url{http://arxiv.org/abs/1206.5083}.
\newblock \eprint{1206.5083}.

\bibitem{Mannhart2010}
\bibinfo{author}{Mannhart, J.} \& \bibinfo{author}{Schlom, D.~G.}
\newblock \bibinfo{title}{Oxide interfaces --- an opportunity for electronics}.
\newblock \emph{\bibinfo{journal}{Science}} \textbf{\bibinfo{volume}{327}},
  \bibinfo{pages}{1607--1611} (\bibinfo{year}{2010}).
\newblock
  \urlprefix\url{http://www.sciencemag.org/content/327/5973/1607.abstract}.

\bibitem{Willmott2007}
\bibinfo{author}{Willmott, P.~R.} \emph{et~al.}
\newblock \bibinfo{title}{Structural basis for the conducting interface between
  {LaAlO$_{3}$} and {SrTiO$_{3}$}}.
\newblock \emph{\bibinfo{journal}{Phys. Rev. Lett.}}
  \textbf{\bibinfo{volume}{99}}, \bibinfo{pages}{155502}
  (\bibinfo{year}{2007}).

\bibitem{Thiel2006}
\bibinfo{author}{Thiel, S.}, \bibinfo{author}{Hammerl, G.},
  \bibinfo{author}{Schmehl, A.}, \bibinfo{author}{Schneider, C.~W.} \&
  \bibinfo{author}{Mannhart, J.}
\newblock \bibinfo{title}{Tunable quasi-two-dimensional electron gases in oxide
  heterostructures}.
\newblock \emph{\bibinfo{journal}{Science}} \textbf{\bibinfo{volume}{313}},
  \bibinfo{pages}{1942--1945} (\bibinfo{year}{2006}).
\newblock
  \urlprefix\url{http://www.sciencemag.org/content/313/5795/1942.abstract}.

\bibitem{Nakagawa2006}
\bibinfo{author}{Nakagawa, N.}, \bibinfo{author}{Hwang, H.~Y.} \&
  \bibinfo{author}{Muller, D.~A.}
\newblock \bibinfo{title}{Why some interfaces cannot be sharp}.
\newblock \emph{\bibinfo{journal}{Nature Mater.}} \textbf{\bibinfo{volume}{5}},
  \bibinfo{pages}{204--209} (\bibinfo{year}{2006}).
\newblock \urlprefix\url{http://dx.doi.org/10.1038/nmat1569}.

\bibitem{Schlom2011}
\bibinfo{author}{Schlom, D.~G.} \& \bibinfo{author}{Mannhart, J.}
\newblock \bibinfo{title}{Oxide electronics: {Interface} takes charge over
  {Si}}.
\newblock \emph{\bibinfo{journal}{Nature Mater.}}
  \textbf{\bibinfo{volume}{10}}, \bibinfo{pages}{168--169}
  (\bibinfo{year}{2011}).
\newblock \urlprefix\url{http://dx.doi.org/10.1038/nmat2965}.

\bibitem{Annadi2012a}
\bibinfo{author}{{Annadi}, A.} \emph{et~al.}
\newblock \bibinfo{title}{{Unexpected Anisotropic Two Dimensional Electron Gas
  at the LaAlO3/SrTiO3 (110) Interface}}.
\newblock \emph{\bibinfo{journal}{ArXiv e-prints}}  (\bibinfo{year}{2012}).
\newblock \urlprefix\url{http://arxiv.org/abs/1208.6135}.
\newblock \eprint{1208.6135}.

\bibitem{Chen2011b}
\bibinfo{author}{Chen, Y.} \emph{et~al.}
\newblock \bibinfo{title}{Metallic and insulating interfaces of amorphous
  {SrTiO$_3$}-based oxide heterostructures}.
\newblock \emph{\bibinfo{journal}{Nano Letters}} \textbf{\bibinfo{volume}{11}},
  \bibinfo{pages}{3774--3778} (\bibinfo{year}{2011}).
\newblock \urlprefix\url{http://pubs.acs.org/doi/abs/10.1021/nl201821j}.
\newblock \eprint{http://pubs.acs.org/doi/pdf/10.1021/nl201821j}.

\bibitem{Chen2013}
\bibinfo{author}{Chen, Y.~Z.} \emph{et~al.}
\newblock \bibinfo{title}{A high-mobility two-dimensional electron gas at the
  spinel/perovskite interface of {$\gamma$-Al$_2$O$_3$/SrTiO$_3$}}.
\newblock \emph{\bibinfo{journal}{Nat. Commun.}} \textbf{\bibinfo{volume}{4}},
  \bibinfo{pages}{1371} (\bibinfo{year}{2013}).
\newblock \urlprefix\url{http://dx.doi.org/10.1038/ncomms2394}.

\bibitem{Herranz2012}
\bibinfo{author}{Herranz, G.}, \bibinfo{author}{S\'anchez, F.},
  \bibinfo{author}{Dix, N.}, \bibinfo{author}{Scigaj, M.} \&
  \bibinfo{author}{Fontcuberta, J.}
\newblock \bibinfo{title}{High mobility conduction at (110) and (111)
  {LaAlO$_3$/SrTiO$_3$} interfaces}.
\newblock \emph{\bibinfo{journal}{Sci. Rep.}} \textbf{\bibinfo{volume}{2}},
  \bibinfo{pages}{758} (\bibinfo{year}{2012}).
\newblock \urlprefix\url{http://dx.doi.org/10.1038/srep00758}.

\bibitem{Plumb2013}
\bibinfo{author}{{Plumb}, N.~C.} \emph{et~al.}
\newblock \bibinfo{title}{{Mixed dimensionality of confined conducting
  electrons tied to ferroelectric surface distortion on an oxide}}.
\newblock \emph{\bibinfo{journal}{ArXiv e-prints}}  (\bibinfo{year}{2013}).
\newblock \urlprefix\url{http://arxiv.org/abs/1302.0708}.
\newblock \eprint{1302.0708}.

\bibitem{Sing2009}
\bibinfo{author}{Sing, M.} \emph{et~al.}
\newblock \bibinfo{title}{Profiling the interface electron gas of
  {LaAlO$_{3}$/SrTiO$_{3}$} heterostructures with hard x-ray photoelectron
  spectroscopy}.
\newblock \emph{\bibinfo{journal}{Phys. Rev. Lett.}}
  \textbf{\bibinfo{volume}{102}}, \bibinfo{pages}{176805}
  (\bibinfo{year}{2009}).

\bibitem{Cancellieri2013}
\bibinfo{author}{Cancellieri, C.} \emph{et~al.}
\newblock \bibinfo{title}{Interface fermi states of
  {${\mathrm{LaAlO}}_{3}/{\mathrm{SrTiO}}_{3}$} and related heterostructures}.
\newblock \emph{\bibinfo{journal}{Phys. Rev. Lett.}}
  \textbf{\bibinfo{volume}{110}}, \bibinfo{pages}{137601}
  (\bibinfo{year}{2013}).
\newblock
  \urlprefix\url{http://link.aps.org/doi/10.1103/PhysRevLett.110.137601}.

\bibitem{Zhou2011}
\bibinfo{author}{Zhou, K.-J.} \emph{et~al.}
\newblock \bibinfo{title}{Localized and delocalized {Ti 3$d$} carriers in
  {LaAlO$_3$/SrTiO$_3$} superlattices revealed by resonant inelastic x-ray
  scattering}.
\newblock \emph{\bibinfo{journal}{Phys. Rev. B}} \textbf{\bibinfo{volume}{83}},
  \bibinfo{pages}{201402(R)} (\bibinfo{year}{2011}).

\bibitem{Pauli2011}
\bibinfo{author}{Pauli, S.~A.} \emph{et~al.}
\newblock \bibinfo{title}{Evolution of the interfacial structure of
  {LaAlO$_{3}$} on {SrTiO$_{3}$}}.
\newblock \emph{\bibinfo{journal}{Phys. Rev. Lett.}}
  \textbf{\bibinfo{volume}{106}}, \bibinfo{pages}{036101}
  (\bibinfo{year}{2011}).
\newblock
  \urlprefix\url{http://link.aps.org/doi/10.1103/PhysRevLett.106.036101}.

\bibitem{Cantoni2012}
\bibinfo{author}{Cantoni, C.} \emph{et~al.}
\newblock \bibinfo{title}{Electron transfer and ionic displacements at the
  origin of the {2D} electron gas at the {LAO/STO} interface: {Direct}
  measurements with atomic-column spatial resolution}.
\newblock \emph{\bibinfo{journal}{Adv. Mater.}} \textbf{\bibinfo{volume}{24}},
  \bibinfo{pages}{3952?--3957} (\bibinfo{year}{2012}).
\newblock \urlprefix\url{http://dx.doi.org/10.1002/adma.201200667}.

\bibitem{Bickel1989}
\bibinfo{author}{Bickel, N.}, \bibinfo{author}{Schmidt, G.},
  \bibinfo{author}{Heinz, K.} \& \bibinfo{author}{M\"uller, K.}
\newblock \bibinfo{title}{Ferroelectric relaxation of the {SrTiO$_{3}$(100)}
  surface}.
\newblock \emph{\bibinfo{journal}{Phys. Rev. Lett.}}
  \textbf{\bibinfo{volume}{62}}, \bibinfo{pages}{2009--2011}
  (\bibinfo{year}{1989}).
\newblock \urlprefix\url{http://link.aps.org/doi/10.1103/PhysRevLett.62.2009}.

\bibitem{Hikita1993}
\bibinfo{author}{Hikita, T.}, \bibinfo{author}{Hanada, T.},
  \bibinfo{author}{Kudo, M.} \& \bibinfo{author}{Kawai, M.}
\newblock \bibinfo{title}{Structure and electronic state of the {TiO$_2$} and
  {SrO} terminated {SrTiO$_3$(100)} surfaces}.
\newblock \emph{\bibinfo{journal}{Surf. Sci.}}
  \textbf{\bibinfo{volume}{287--288}}, \bibinfo{pages}{377--381}
  (\bibinfo{year}{1993}).
\newblock
  \urlprefix\url{http://www.sciencedirect.com/science/article/pii/003960289390806U}.

\bibitem{Ikeda1999}
\bibinfo{author}{Ikeda, A.}, \bibinfo{author}{Nishimura, T.},
  \bibinfo{author}{Morishita, T.} \& \bibinfo{author}{Kido, Y.}
\newblock \bibinfo{title}{Surface relaxation and rumpling of
  {TiO$_2$}-terminated {SrTiO$_3(001)$} determined by medium energy ion
  scattering}.
\newblock \emph{\bibinfo{journal}{Surf. Sci.}}
  \textbf{\bibinfo{volume}{433--435}}, \bibinfo{pages}{520--524}
  (\bibinfo{year}{1999}).
\newblock
  \urlprefix\url{http://www.sciencedirect.com/science/article/pii/S0039602899000503}.

\bibitem{Popovic2008}
\bibinfo{author}{Popovi\'{c}, Z.~S.}, \bibinfo{author}{Satpathy, S.} \&
  \bibinfo{author}{Martin, R.~M.}
\newblock \bibinfo{title}{Origin of the two-dimensional electron gas carrier
  density at the {LaAlO$_{3}$} on {SrTiO$_{3}$} interface}.
\newblock \emph{\bibinfo{journal}{Phys. Rev. Lett.}}
  \textbf{\bibinfo{volume}{101}}, \bibinfo{pages}{256801}
  (\bibinfo{year}{2008}).

\bibitem{Copie2009}
\bibinfo{author}{Copie, O.} \emph{et~al.}
\newblock \bibinfo{title}{Towards two-dimensional metallic behavior at
  {LaAlO$_{3}$/SrTiO$_{3}$} interfaces}.
\newblock \emph{\bibinfo{journal}{Phys. Rev. Lett.}}
  \textbf{\bibinfo{volume}{102}}, \bibinfo{pages}{216804}
  (\bibinfo{year}{2009}).
\newblock
  \urlprefix\url{http://link.aps.org/doi/10.1103/PhysRevLett.102.216804}.

\bibitem{Dubroka2010}
\bibinfo{author}{Dubroka, A.} \emph{et~al.}
\newblock \bibinfo{title}{Dynamical response and confinement of the electrons
  at the {LaAlO$_{3}$/SrTiO$_{3}$} interface}.
\newblock \emph{\bibinfo{journal}{Phys. Rev. Lett.}}
  \textbf{\bibinfo{volume}{104}}, \bibinfo{pages}{156807}
  (\bibinfo{year}{2010}).
\newblock
  \urlprefix\url{http://link.aps.org/doi/10.1103/PhysRevLett.104.156807}.

\bibitem{Delugas2011}
\bibinfo{author}{Delugas, P.} \emph{et~al.}
\newblock \bibinfo{title}{Spontaneous 2-dimensional carrier confinement at the
  $n$-type {SrTiO$_{3}$/LaAlO$_{3}$} interface}.
\newblock \emph{\bibinfo{journal}{Phys. Rev. Lett.}}
  \textbf{\bibinfo{volume}{106}}, \bibinfo{pages}{166807}
  (\bibinfo{year}{2011}).

\bibitem{Stengel2011}
\bibinfo{author}{Stengel, M.}
\newblock \bibinfo{title}{First-principles modeling of electrostatically doped
  perovskite systems}.
\newblock \emph{\bibinfo{journal}{Phys. Rev. Lett.}}
  \textbf{\bibinfo{volume}{106}}, \bibinfo{pages}{136803}
  (\bibinfo{year}{2011}).
\newblock
  \urlprefix\url{http://link.aps.org/doi/10.1103/PhysRevLett.106.136803}.

\bibitem{Savoia2009}
\bibinfo{author}{Savoia, A.} \emph{et~al.}
\newblock \bibinfo{title}{Polar catastrophe and electronic reconstructions at
  the {LaAlO$_{3}$/SrTiO$_{3}$} interface: {Evidence} from optical second
  harmonic generation}.
\newblock \emph{\bibinfo{journal}{Phys. Rev. B}} \textbf{\bibinfo{volume}{80}},
  \bibinfo{pages}{075110} (\bibinfo{year}{2009}).
\newblock \urlprefix\url{http://link.aps.org/doi/10.1103/PhysRevB.80.075110}.

\bibitem{BenShalom2010}
\bibinfo{author}{Ben~Shalom, M.}, \bibinfo{author}{Ron, A.},
  \bibinfo{author}{Palevski, A.} \& \bibinfo{author}{Dagan, Y.}
\newblock \bibinfo{title}{{Shubnikov-De Haas} oscillations in
  {SrTiO$_{3}$/LaAlO$_{3}$} interface}.
\newblock \emph{\bibinfo{journal}{Phys. Rev. Lett.}}
  \textbf{\bibinfo{volume}{105}}, \bibinfo{pages}{206401}
  (\bibinfo{year}{2010}).
\newblock
  \urlprefix\url{http://link.aps.org/doi/10.1103/PhysRevLett.105.206401}.

\bibitem{Koitzsch2011a}
\bibinfo{author}{Koitzsch, A.} \emph{et~al.}
\newblock \bibinfo{title}{In-gap electronic structure of {LaAlO$_3$-SrTiO$_3$}
  heterointerfaces investigated by soft x-ray spectroscopy}.
\newblock \emph{\bibinfo{journal}{Phys. Rev. B}} \textbf{\bibinfo{volume}{84}},
  \bibinfo{pages}{245121} (\bibinfo{year}{2011}).

\bibitem{Drera2011}
\bibinfo{author}{Drera, G.} \emph{et~al.}
\newblock \bibinfo{title}{Spectroscopic evidence of in-gap states at the
  {SrTiO$_3$/LaAlO$_3$} ultrathin interfaces}.
\newblock \emph{\bibinfo{journal}{Appl. Phys. Lett.}}
  \textbf{\bibinfo{volume}{98}}, \bibinfo{pages}{052907}
  (\bibinfo{year}{2011}).
\newblock \urlprefix\url{http://link.aip.org/link/?APL/98/052907/1}.

\bibitem{Abbate1991}
\bibinfo{author}{Abbate, M.} \emph{et~al.}
\newblock \bibinfo{title}{Soft-x-ray-absorption studies of the location of
  extra charges induced by substitution in controlled-valence materials}.
\newblock \emph{\bibinfo{journal}{Phys. Rev. B}} \textbf{\bibinfo{volume}{44}},
  \bibinfo{pages}{5419--5422} (\bibinfo{year}{1991}).
\newblock \urlprefix\url{http://link.aps.org/doi/10.1103/PhysRevB.44.5419}.

\bibitem{Salluzzo2009}
\bibinfo{author}{Salluzzo, M.} \emph{et~al.}
\newblock \bibinfo{title}{Orbital reconstruction and the two-dimensional
  electron gas at the {LaAlO$_{3}$/SrTiO$_{3}$} interface}.
\newblock \emph{\bibinfo{journal}{Phys. Rev. Lett.}}
  \textbf{\bibinfo{volume}{102}}, \bibinfo{pages}{166804}
  (\bibinfo{year}{2009}).

\bibitem{Zhong2008}
\bibinfo{author}{Zhong, Z.} \& \bibinfo{author}{Kelly, P.~J.}
\newblock \bibinfo{title}{Electronic-structure-induced reconstruction and
  magnetic ordering at the {LaAlO$_3$|SrTiO$_3$} interface}.
\newblock \emph{\bibinfo{journal}{Europhys. Lett.}}
  \textbf{\bibinfo{volume}{84}}, \bibinfo{pages}{27001} (\bibinfo{year}{2008}).
\newblock \urlprefix\url{http://stacks.iop.org/0295-5075/84/i=2/a=27001}.

\bibitem{Segal2009}
\bibinfo{author}{Segal, Y.}, \bibinfo{author}{Ngai, J.~H.},
  \bibinfo{author}{Reiner, J.~W.}, \bibinfo{author}{Walker, F.~J.} \&
  \bibinfo{author}{Ahn, C.~H.}
\newblock \bibinfo{title}{X-ray photoemission studies of the metal-insulator
  transition in {${\text{LaAlO}}_{3}/{\text{SrTiO}}_{3}$} structures grown by
  molecular beam epitaxy}.
\newblock \emph{\bibinfo{journal}{Phys. Rev. B}} \textbf{\bibinfo{volume}{80}},
  \bibinfo{pages}{241107} (\bibinfo{year}{2009}).
\newblock \urlprefix\url{http://link.aps.org/doi/10.1103/PhysRevB.80.241107}.

\bibitem{Slooten2013}
\bibinfo{author}{{Slooten}, E.} \emph{et~al.}
\newblock \bibinfo{title}{{Hard x-ray photoemission and density functional
  theory study of the internal electric field in SrTiO3/LaAlO3 oxide
  heterostructures}}.
\newblock \emph{\bibinfo{journal}{ArXiv e-prints}}  (\bibinfo{year}{2013}).
\newblock \urlprefix\url{http://arxiv.org/abs/1301.2179}.
\newblock \eprint{1301.2179}.

\bibitem{Berner2013}
\bibinfo{author}{{Berner}, G.} \emph{et~al.}
\newblock \bibinfo{title}{{Direct k-space mapping of the electronic structure
  in an oxide-oxide interface}}.
\newblock \emph{\bibinfo{journal}{ArXiv e-prints}}  (\bibinfo{year}{2013}).
\newblock \urlprefix\url{http://arxiv.org/abs/1301.2824}.
\newblock \eprint{1301.2824}.

\bibitem{Basletic2008}
\bibinfo{author}{Basletic, M.} \emph{et~al.}
\newblock \bibinfo{title}{Mapping the spatial distribution of charge carriers
  in {LaAlO$_3$/SrTiO$_3$} heterostructures}.
\newblock \emph{\bibinfo{journal}{Nature Mater.}} \textbf{\bibinfo{volume}{7}},
  \bibinfo{pages}{621--625} (\bibinfo{year}{2008}).
\newblock \urlprefix\url{http://dx.doi.org/10.1038/nmat2223}.

\bibitem{Schwingenschlogl2009}
\bibinfo{author}{Schwingenschl\"ogl, U.} \& \bibinfo{author}{Schuster, C.}
\newblock \bibinfo{title}{Interface relaxation and electrostatic charge
  depletion in the oxide heterostructure {LaAlO$_3$/SrTiO$_3$}}.
\newblock \emph{\bibinfo{journal}{Europhys. Lett.}}
  \textbf{\bibinfo{volume}{86}}, \bibinfo{pages}{27005} (\bibinfo{year}{2009}).
\newblock \urlprefix\url{http://stacks.iop.org/0295-5075/86/i=2/a=27005}.

\bibitem{Stengel2012}
\bibinfo{author}{Stengel, M.}, \bibinfo{author}{Fennie, C.~J.} \&
  \bibinfo{author}{Ghosez, P.}
\newblock \bibinfo{title}{Electrical properties of improper ferroelectrics from
  first principles}.
\newblock \emph{\bibinfo{journal}{Phys. Rev. B}} \textbf{\bibinfo{volume}{86}},
  \bibinfo{pages}{094112} (\bibinfo{year}{2012}).
\newblock \urlprefix\url{http://link.aps.org/doi/10.1103/PhysRevB.86.094112}.

\bibitem{Bousquet2008}
\bibinfo{author}{Bousquet, E.} \emph{et~al.}
\newblock \bibinfo{title}{Improper ferroelectricity in perovskite oxide
  artificial superlattices}.
\newblock \emph{\bibinfo{journal}{Nature}} \textbf{\bibinfo{volume}{452}},
  \bibinfo{pages}{732--736} (\bibinfo{year}{2008}).
\newblock \urlprefix\url{http://dx.doi.org/10.1038/nature06817}.

\bibitem{Morozovska2012}
\bibinfo{author}{Morozovska, A.~N.}, \bibinfo{author}{Eliseev, E.~A.},
  \bibinfo{author}{Glinchuk, M.~D.}, \bibinfo{author}{Chen, L.-Q.} \&
  \bibinfo{author}{Gopalan, V.}
\newblock \bibinfo{title}{Interfacial polarization and pyroelectricity in
  antiferrodistortive structures induced by a flexoelectric effect and
  rotostriction}.
\newblock \emph{\bibinfo{journal}{Phys. Rev. B}} \textbf{\bibinfo{volume}{85}},
  \bibinfo{pages}{094107} (\bibinfo{year}{2012}).
\newblock \urlprefix\url{http://link.aps.org/doi/10.1103/PhysRevB.85.094107}.

\bibitem{Salluzzo2013}
\bibinfo{author}{Salluzzo, M.} \emph{et~al.}
\newblock \bibinfo{title}{Structural and electronic reconstructions at the
  {LaAlO$_3$/SrTiO$_3$} interface}.
\newblock \emph{\bibinfo{journal}{Adv. Mater.}}  (\bibinfo{year}{2013}).
\newblock \urlprefix\url{http://dx.doi.org/10.1002/adma.201204555}.

\end{thebibliography}

\end{document}